\documentstyle[aps,preprint]{revtex}

\begin{document}
\title{
Static displacements and chemical correlations in alloys}
%\draft
\author{Bulbul Chakraborty}
\address{
The Martin Fisher School of Physics\\
Brandeis University\\
Waltham, MA 02254, USA}
%\date{\today}
\maketitle

\begin{abstract}
Recent experiments in metallic solid solutions
have revealed interesting correlations between static pair-displacements
and the ordering behavior of these alloys.
This paper discusses a simple theoretical model
which successfully
explains these observations and which provides a natural framework
for analyzing
experimental measurements of pair-displacements and chemical
correlations in solid solutions.   The utility and
scope of this model is demonstrated by analyzing results of
experiments on $Ni-Fe$ and $Cr-Fe$ alloys and results of
simulations of $Cu-Au$ and $Cu-Ag$ alloys.
\end{abstract}
\narrowtext

Metallic alloys which exist as crystalline solid solutions at high
temperature are classified, according to their low-temperature
behavior, as ordering or clustering alloys.
%can be divided into two categories; those which exhibit a
%low-temperature ordered phase and those which cluster or phase separate.
Investigation
of the local correlations in the high-temperature phase yields important
information regarding the low-temperature behavior and is an extensively
used
tool in the study of alloy phase stability.  Measurements of chemical
correlations have been used to extract effective interaction parameters
in theoretical models\cite{prlsparks}.  In recent experiments, the
measurements were extended to include chemically specific {\it
displacements} of atoms\cite{prlsparks,natosparks}.
A striking observation of these experiments
was a correlation between $AB$ distances and the clustering or ordering
tendency of the $A-B$ alloy\cite{natosparks}.  It was
observed that the $AB$ distances were the shortest distances in an
ordering alloy; contrary to expectations based on "size-effect"
arguments\cite{prlsparks,natosparks}.

The purpose of this
paper is to present a theoretical model which successfully explains the
correlation between pair-displacements and the ordering behavior of
alloys.
The most important predictions of this model are: (i) the
$AB$ distances are always shorter than the arithmetic
mean of the $AA$ and $BB$ distances in an ordering alloy and larger than
the arithmetic mean in a clustering alloy and (ii) the $AB$ distances
are the {\it shortest} in an ordering alloy where the size difference is
negligible.  This model is the first of its kind to lead to these
predictions, especially the second, and identifies the types
of interactions needed to understand the experimental observations.
Simulations of alloys with large size differences follow the trend of
prediction (i) as demonstrated by embedded atom
studies\cite{thorpe,foot0}.

The model under discussion is a compressible Ising model\cite{kardar},
extended to include size-effect terms which are non-existent in a
magnetic system but can be extremely important in binary alloys.   This
model is referred to as the generalized Ising model in the following.
The compressible Ising model provides a simple description of
order-disorder transitions in alloys which are accompanied by a
displacive structural transition\cite{natoemt,leigu}.  An example is the
ordering transition in $CuAu$ which is accompanied by a cubic to
tetragonal distortion\cite{natoemt,emt1}.
The present work shows that this
model provides a good description of displacements in
binary alloys including
transition metal alloys whose microscopic interactions are very different
from those of $Cu-Au$ alloys.  The generalized Ising model differs from
the model usually used in studying displacements in alloys through the
inclusion of a distance-dependent Ising interaction.

The issue of displacements in alloys has been addressed extensively in
the literature\cite{cook,froyen,zunger}.  More recently, first
principles electronic structure calculations have also been used to
obtain the relevant parameters for the model Hamiltonian defining the
coupling between displacements and configuration degrees of
freedom\cite{zunger,gironcoli}.
The variables in such a model are $\{ {\bf u}_i \}$, the
displacements at site $i$, and the occupation or Ising variables, $\{ S_i
\} =\pm  1$.
A displaced Ising variable defined
by $S_i = S_i - <S>$ can be used, where $<S>$ measures the average
concentration and is zero for a 50-50 alloy.
In terms of these variables,
the standard model describing
binary alloys can be written as\cite{cook,gironcoli}
\begin{equation}
\label{model}
 H = {\sum}_{i,j} [ J_{ij} S_i S_j + {\sum}_{\alpha  \beta}
u_{i{\alpha}}D_{{\alpha}{\beta}} (i,j) u_{j{\beta}} +{\sum}_{\alpha}
K_{\alpha} (i,j) u_{i{\alpha}}S_j ]
\end{equation}
In this equation, the second term is the elastic energy and the third
term describes the {\it only} coupling between the displacements and the
spins.  The $K_{ij}$ are the forces induced due to the difference
between the two types of atoms\cite{cook,gironcoli} and the $J_{ij}$ are
the pair interactions.
This model has been used to analyze pair displacements in
alloys\cite{froyen}, and it has been shown that the average pair
displacement of sites $m$ and $n$ depend on $<S_{m}> + <S_{n}>$. It
follows immediately that this model would always predict an $AB$
distance which is {\it intermediate} between the $AA$ and $BB$
distances; never the shortest or longest.
Another feature of this model is that the average
separations do not depend on the configurational (chemical)
short-range order (the {\it mean-square} separations do\cite{froyen})
and there is no direct correlation between the
average
separations and the ordering or clustering tendency of the
solid solution.  In contrast, the analysis below will show that such a
correlation comes about naturally in the generalized Ising model.

The generalized Ising model
differs in one crucial respect from the above model for alloys.  The
difference lies in the displacement-dependence of the Ising interaction
$J_{ij}$; characteristic of a compressible Ising model:
\begin{equation}
\label{compress}
J_{ij} = J^{0}_{ij}(1 - {\epsilon}({\bf u}_i - {\bf
u}_j)\cdot{{\bf R}_{ij}/|{\bf R}_{ij}|}) ~.
\end{equation}
Here, the ${\bf R}_{ij}$ are the distances of the undistorted
lattice, and the parameter ${\epsilon}$ determines the strength of the
coupling to the lattice.
The new feature that this introduces into the alloy model is the
coupling of displacements to the {\it relative} spin orientation, $S_i
S_j$ which naturally leads to a relationship between static
displacements and configurational order.  It is clear that the pair
displacements predicted by the model defined by
Eqs. (\ref{model}) and (\ref{compress})
will depend on the sign of the Ising interaction\cite{footising}.

The physics embodied in the compressible Ising model is very simple, it
expresses the fact that the interaction between different types
of atoms, whether
attractive or repulsive, depends on the distance between them.
The arguments leading to the
linear dependence on displacements in Eq. (\ref{compress}) are very
general;
(i) it is the leading term in any
expansion and should be adequate for small displacements\cite{kardar} (ii)
it would be obtained in an analysis of the type carried out in Cook and
de Fontaine\cite{cook} or in Gironcoli {\it et al}\cite{gironcoli} if the
expansion retains terms which are linear in displacements and quadratic
in the spins and, (iii) moreover, it has a  microscopic basis in
metallic alloys
arrived at through the analysis of the ordering behavior in
$Cu-Au$ alloys\cite{natoemt,emt1}.

Since the interactions depend on
the distance in the generalized Ising model,
it seems natural that static displacements accompany
chemical correlations.  A striking demonstration  of this is seen in the
triangular lattice antiferromagnet\cite{kardar} and in its alloy
analog, the $L1_0$
structure on a face-centered-cubic lattice.  The ordered phases here
have the ferromagnetic (like-atom) bonds elongated and the
antiferromagnetic (unlike-atom) bonds shortened\cite{natoemt,emt1}.
It would not
be a surprise, therefore, to find such correlations in the high temperature
disordered phase of an alloy.

It should be emphasized that the generalized Ising model is designed to
describe the basic types of interactions in an alloy and the
parameters relevant to a particular alloy can be obtained from {\it
ab initio}
electronic structure calculations or from embedded atom type
approaches\cite{emtrev,dawandbaskes}.
All the parameters in the model are expected to depend on the type and
concentration of alloys.
The object of this work is to show that the
distance-dependence of the Ising interaction leads to qualitatively
different behavior of the pair displacements and that these predictions
can explain experimental observations.

In the following discussion,
it will be assumed
that the displacements are being measured with respect to the  average
lattice including global distortions, wherever they are present.
Mechanical equilibrium implies that for a given configuration of spins
(chemical configuration of the alloy), the static displacements are
those obtained by minimizing the energy.
The displacements for a
given thermodynamic state can then be obtained by thermally averaging
over the spin configurations.

Minimizing the energy given in Eq. (\ref{model}) with respect to the
displacements
${\bf u}_i$, and carrying out the statistical averaging
leads to the
following expression:
\begin{eqnarray}
\label{displac}
<{\bf u}_m - {\bf u}_n >& = & {\sum}_{\bf q} \{ \exp (i{\bf q}\cdot{\bf R}_m )
- \exp (-i{\bf q}\cdot{\bf R}_n )\} \nonumber\\
& {\sum}_{ij} & \exp
(-i {\bf q}\cdot{\bf R}_i )  D^{-1}_{\bf q} ({{\bf R}_{ij}/|{\bf
R}_{ij}|}) [ J^{0}_{ij} {\epsilon} <S_i S_j >+ K_{ij} <S_j >]
\end{eqnarray}

The size-effect term, involving $K_{ij}$, is of the exact same form as
Froyen and
Herring with a cubic symmetry assumed implying
that the forces are along the radial direction\cite{footnot1}.  This
facilitates the comparison between the two models.  The elastic
properties of the lattice appear in
Eq. (\ref{displac}) through the
dynamical matrix,
$D_{\bf q}$.  All the interaction parameters in the model Hamitonian,
Eq. (\ref{model}), are bare interaction parameters which do not depend
on the displacements or the spin configurations.  The statistical
averages in Eq. (\ref{displac}) are taken with the complete Hamiltonian
defined by Eqs. (\ref{model}) and (\ref{compress}) and include
renormalization of these bare parameters which occur naturally in these
models.  As an example, the effective elastic moduli depend on
temperature\cite{kardar}.  In obtaining parameters from microscopic
calculations, the natural choice is the parameters of the random solid
solution.

It is clear that the Ising interaction and the size-effect term
play different roles in determining the static displacements.  The
crucial difference is that the distance-dependent Ising term
leads to a static displacement
proportional to the pair-correlation function, the Warren-Cowley
short-range order parameter\cite{prlsparks}.
Analysis of the simplest situation,
a random 50-50
alloy with only nearest neighbor interactions,
demonstrates these differences.  Experiments measure
the relative displacements,
$u^{\mu \nu}_{mn} =
<{\bf u}_m - {\bf u}_n>^{\mu \nu} $,
for a ${\mu \nu}$ pair situated at sites $m$
and $n$ where the ${\mu \nu}$
can be $AA$, $BB$,or $AB$ for an $A-B$ alloy.

For a random alloy, all the averages in Eq. (\ref {displac}) are zero
except the ones involving sites $m$ and $n$.  For this situation, Eq.
(\ref {displac}) predicts
\begin{equation}
\label{dmodel}
u^{\mu \nu}_{mn} = 2J^{0}\epsilon G_{mn} <S_m S_n>_{\mu \nu} + K I_{mn}
(<S_m>_{\mu \nu} +
<S_n>_{\mu \nu}) ~ ,
\end{equation}
where
$$
G_{mn} = {\sum}_{\bf q}  D^{-1}_{\bf q} ({{\bf R}_{mn} /|{\bf
R}_{mn}|}) (1- \cos {\bf q}\cdot{\bf R}_{mn} ) ~,
$$
$K$ corresponds to $K_{ij}$ where $i$ and $j$ are nearest neighbors,
and $I_{mn}$ is exactly the same as that defined in Eq. 20 of
Froyen-Herring\cite{froyen}.
The values of the averages for different pairs in the 50-50 alloy are,
$<S_m S_n>_{AA} = <S_m S_n>_{BB} = 1$, $<S_m S_n>_{AB} = -1$, and
$(<S>_m + <S>_n)_{AA} = 2$,
$(<S>_m + <S>_n)_{BB} = -2$,
$(<S>_m + <S>_n)_{AB} = 0$.

The size effect term, involving K, leads to a zero relative displacement of
$A$ and $B$ atom-pairs in a 50-50 random alloy whereas the
Ising term leads to a non-zero static displacement.
The results of simulations\cite{thorpe} show that
the average $AB$ distances can be quite different from the average
lattice spacing in a 50-50 alloy.
The simulations of
$Cu_{50}Au_{50}$ and $Cu_{50}Ag_{50}$, to be discussed below,
also show a nonzero displacement of $AB$ pairs.

Experimental measurements in $Ni_{77.5}Fe_{22.5}$, an ordering alloy,
show that the $Ni-Fe$ nearest-neighbor pairs are the closest neighbors
in the solid solution.  In $Cr_{47}Fe_{53}$, a clustering alloy, the
experiments find that the $Cr-Fe$ nearest-neighbor pairs are the
farthest apart.  These observations are explained naturally by the
generalized Ising model if it is assumed that the Ising term is more
important than the size-effect term.  For these alloys, this is not an
unreasonable assumption since the sizes are similar.
The experimental values of the nearest-neighbor displacements of
the $AB$ pairs in the $NiFe$  and $CrFe$ alloys can be fitted,
successfully,  to the
expression for the displacements, Eq. (\ref{displac}).
To
simplify the analysis, the short-range order was neglected and the
alloys were assumed to be random.  The inclusion of the short-range
order, which is not very large in the experimental systems,
would change the quantitative values of the interaction parameters
but will not affect the qualitative behavior.
The parameters obtained from the fitting imply an
antiferromagnetic interaction
for $Ni_{77.5}Fe_{22.5}$ and a ferromagnetic interaction
for $Cr_{47}Fe_{53}$.
Specifically, the combination
$J^{0}\epsilon G_{mn}$ is found to be equal to 0.071\AA\
for the $Ni-Fe$ alloy
and $-$0.151\AA\ for the $Cr-Fe$ alloy.  The term originating from
size effect, $K I_{mn}$, obtained by fitting the experiments to Eq. 3,
is $-$0.0204\AA\ for the
$Ni-Fe$ and 0.0097\AA\ for the $Cr-Fe$ alloy reflecting the fact that Ni is a
smaller atom than Fe but Cr is larger than Fe.  It is also reassuring to
find a smaller size-effect term for $Cr-Fe$ than for $Ni-F$e since the
constituents of the former have a smaller size difference.

It should be stressed that without  the distance-dependent Ising term, the
experimental data cannot be fit at all to the expression, Eq.
(\ref{displac}), for the static
displacement
\cite{prlsparks,natosparks}.

The alloys studied experimentally, $Ni-Fe$ and $Cr-Fe$, are fairly complicated
systems, being transition metal alloys where magnetism is expected to
play a role in the ordering.  Microscopic models of these alloys,
especially their pair correlation functions, are based on the KKR-CPA
approach\cite{cpareview}.  This approach has been generalized recently
to include displacements of atoms\cite{pinski}.    It
would be interesting to see if the results of this microscopic
approach lead to a description akin to the generalized Ising model
presented here and whether the interaction parameters obtained in the
by fitting to experiment can
be explained on the basis of microscopic interactions.

The connection with the work of Froyen and Herring is
evident
when the general expression for the displacements is simplied further.
To achieve this, two functions need to be
defined:
$$
F({\bf q}) = {\sum}_{(ij)} J^{0}_{ij} \epsilon <S_i S_j> \exp (i{\bf
q}\cdot{\bf
R}_{ij})
$$
and
$$
L({\bf q}) = {\sum}_{(ij)} \exp (i{\bf q}\cdot{\bf R}_{ij})
$$
In terms of these, the expression for the displacements is,
\begin{eqnarray}
\label{disp1}
<{\bf u}_m - {\bf u}_n >& = & {\sum}_{\bf q} D^{-1}_{\bf q} ({\nabla}_{\bf
q} F({\bf q})) C^{\prime}_{mn}\nonumber\\
& + & {\sum}_{\bf q} K D^{-1}_{\bf q}  ({\nabla}_{\bf q} L({\bf q}))
C^{\prime \prime}_{mn} ~.
\end{eqnarray}
Here, $C^{\prime}_{mn} = {\sum}_{j}[\exp (i{\bf q})\cdot{\bf R}_{jm} -
\exp (i{\bf q}\cdot{\bf R}_{jn})]$
and $C^{\prime \prime}_{mn}$ has a similar definition with the factor in the
summation being multiplied by $<S_j>$.  Again, in keeping with the work
of Froyen and Herring, the size effect term $K_{ij}$ has been assumed to
be short-range and $L({\bf q})$ involves only a sum over nearest
neighbor shells.
If $\epsilon$ is set to zero, then Eq. (\ref{disp1}) is
identical to Eq. 19 of Froyen and Herring\cite{froyen}.

The two terms in the model that are responsible for the static displacement,
the compressible Ising and the size-effect term, compete with
each other and the $AB$ distance in an ordering alloy does not
necessarily come out to be smaller than the $AA$ distance ($A$ is the
smaller atom).  If K is small, the Ising term dominates and the $AB$
distance indeed turns out to be either the shortest (ordering) or
longest (clustering).  If the coupling to the lattice, $\epsilon$, is
very small then the results from size-effect models are recovered.

One
simple class of microscopic models which can deal with both structural
and
chemical changes is the embedded atom approach which include the
formalism
of Daw and
Baskes \cite{dawandbaskes} and the
Effective Medium Theory (EMT) of cohesion
in metals\cite{emtrev}.
The latter approach has been used to simulate ordering
and kinetics of ordering in $Cu-Au$ alloys and shown to lead to a very
good
description of these phenomena.  In the present work, this
approach was used to obtain pair displacements in alloys with a large size
difference and  the results were compared to the predictions of the generalized
Ising model.

Simulations based on the EMT model\cite{emt1,bcprl}
were performed to obtain chemically specific
average pair displacements in the random solid solutions
$Cu_{50}Au_{50}$ and $Cu_{50}Ag_{50}$.  The simulations were carried out
in a 64 atom cell and the averaging was done over an ensemble of 50
samples.  These simulations allowed for global volume changes besides
individual displacements of atoms.
The results for the
nearest and next-nearest neighbor displacements are shown in Tables
\ref{tab:one} and \ref {tab:two}.

The difference between the ordering and the clustering alloys is clearly
seen although the differences are not as striking
as that between the $Ni-Fe$ and $Cr-Fe$ alloys.
There is increased
competition between the size-effect and the Ising terms in
the simulated alloys arising from the larger size difference between $Cu$
and $Au$ or $Ag$.
Fitting the results of the simulations to the
compressible Ising model leads to a positive value of the
nearest-neighbor $J^0$, {\it i.e.}
antiferromagnetic interactions for  both alloys.  The second neighbor
interactions are different and is antiferromagnetic for $CuAu$\cite{footemt}
but
ferromagnetic for $CuAg$.  The size effect term is larger for the
clustering alloy ({\it cf. Tables \ref{tab:one} and \ref{tab:two}}).

The EMT results for the pair displacements are in qualitative agreement with
the results obtained from the EAM simulations\cite{thorpe}.  Comparing
the two sets of results ({\it cf} Tables I and II),
it is seen that the relative differences
between displacements of the $AA$, $AB$, and $BB$ pairs are very similar
in both methods.  However, the absolute magnitudes of the displacements
are much larger (by nearly a factor of 8)
in the EAM simulations.  The reason for this is not
clear at present.

To conclude, it has been shown that the generalized Ising model
provides a natural
framework for understanding the correlations between static
displacements and chemical short-range order.  In particular, it can
explain the experimental observation that in ordering alloys, the unlike
atoms can be the closest of any nearest-neighbor pair and in clustering
alloys they can be the farthest apart.
The distance-dependence
of the Ising interaction was crucial in obtaining these correlations.
This implies that adding this term to the standard model describing
displacements in alloys would lead to qualitatively new features in the
ordering and phase separating characteristics of alloys.  {\it Ab initio}
electronic structure calculations could be used to obtain the parameters
in the generalized Ising model thus leading to a microscopic
understanding of the interactions.

The simulations were carried out using the ARTwork
simulation package.  The
author would like to acknowledge useful discussions with Frank Pinski
and Cullie Sparks.
This work was supported
in part by DOE through DE-FG02-ER45495.

\begin{table}
\caption{Pair-displacements in a random $Cu_{50}Au_{50}$ alloy obtained
from simulations
based on EMT and the parameters obtained by fitting these displacements
to the generalized
Ising model.
The displacements are measured from the average lattice
and the averages are over an
ensemble consisting of approximately 50 samples. Approximate EAM
results obtained from the plots of Ref. 3
are quoted in parentheses next to the EMT results}
\begin{tabular}{llll}
&&Nearest Neighbor(a. u.)&Next Nearest Neighbor(a. u.)\\
    \tableline
&Cu-Cu&$-$0.0109($\simeq -0.08$)& 0.004\\
&Cu-Au&$-$0.0057($\simeq -0.04$)&$-$0.004\\
&Au-Au& 0.0233($\simeq 0.16$)& 0.005\\
\tableline
&$J^0 \epsilon G_{mn}$(a. u.)& 0.012& 0.016\\
&$K I_{mn}$(a. u.) &$-$0.017& 0.0005\\
\tableline
\end{tabular}
\label{tab:one}
\end{table}

\begin{table}
\caption{Displacements and parameters as in Table 1 but for
the $Cu_{50}Ag_{50}$
alloy.}
\begin{tabular}{llll}
&&Nearest Neighbor(a. u.)&Next Nearest Neighbor(a. u.)\\
    \tableline
&Cu-Cu&$-$0.0144($\simeq -0.12$)&$-$0.018\\
&Cu-Ag&$-$0.007($\simeq 0.01$)& 0.0086\\
&Ag-Ag& 0.0316($\simeq 0.16$)& 0.0036\\
\tableline
&$J^0 \epsilon G_{mn}$(a. u.)& 0.0156&$-$0.0128\\
&$K I_{mn}$(a. u.)&$-$0.023&$-$0.011\\
\tableline
\end{tabular}
\label{tab:two}
\end{table}


\begin{references}
\bibitem{footising} The model represented by Eq. (1) without a
distance-dependent Ising term can be shown to lead to an effective spin
Hamiltonian with a long-range spin-spin interaction.  In contrast, the
distance-dependent Ising term leads to a long-range four spin
interaction (Ref. 5).
\end{references}
\end{document}